# Separation of Electromagnetic and Chemical Contributions to Surface-Enhanced Raman Spectra on Nanoengineered Plasmonic Substrates

*Semion K. Saikin,[1,†] Yizhuo Chu,[2] Dmitrij Rappoport,[1] Kenneth B. Crozier,[2] and Alan Aspuru-Guzik[1,*]*

[1] Department of Chemistry and Chemical Biology, Harvard University, Cambridge, MA 02138, USA

[2] School of Engineering and Applied Sciences, Harvard University, Cambridge, MA 02138, USA

[†] also at the Department of Physics, Kazan State University, Kazan 420008, Russian Federation

ABSTRACT: Raman signals from molecules adsorbed on a noble metal surface are enhanced by many orders of magnitude due to the plasmon resonances of the substrate. Additionally, the enhanced spectra are modified compared to the spectra of neat molecules: many vibrational frequencies are shifted and relative intensities undergo significant changes upon attachment to the metal. With the goal of devising an effective scheme for separating the electromagnetic and chemical effects, we explore the origin of the Raman spectra modification of benzenethiol adsorbed on nanostructured gold surfaces. The spectral modifications are attributed to the frequency dependence of the electromagnetic enhancement and to the effect of chemical binding. The latter contribution can be reproduced computationally using molecule-metal cluster models. We present evidence that the effect of chemical binding is mostly due to changes in the electronic structure of the molecule rather than to the fixed orientation of molecules relative to the substrate.

Surface enhanced Raman spectroscopy [1, 2] provides a highly sensitive molecular detection method with great potential for chemical sensor applications [3]. It can be used for monitoring water pollution [4] or for the detection of biohazards [5]. The phenomenon of surface enhanced Raman scattering (SERS) [6, 7] results in an overall enhancement of molecular Raman response by a factor of often about $10^6$-$10^8$ due to the interaction of the analyte with rough noble metal surfaces [2, 8]. The signal enhancement is attributed to a strong amplification of the electromagnetic fields near the plasmon resonances of metal substrates [2] and to spatially-local effects, which are commonly referred to as chemical enhancement. The characteristic length scale for the electromagnetic enhancement effect is determined by the evanescent nature of localized plasmon modes to be ~10 – 100 nm, and can be significantly modified by surface imperfections. The chemical enhancement effect is characterized by an Ångström length scale and has been associated with metal-molecule electron tunneling [9], intensity borrowing [10], and formation of charge-transfer [11] or metal-molecular surface [12, 13] states. The determination of the absolute value of the chemical enhancement factor in surface-enhanced spectra remains a difficult problem, since measured spectra exhibit the combined effect of both electromagnetic and chemical mechanisms. The inhomogeneous distribution of electromagnetic and chemical enhancement effects, associated with uncontrollable surface imperfections, introduces an additional challenge into the task. In this work, we demonstrate that progress toward understanding chemical enhancement can be made nonetheless by analyzing the relative, rather than absolute, value of chemical enhancement. The method makes use of the plasmon extinction spectrum to approximate the electromagnetic enhancement on nanoengineered substrates, allowing the relative chemical enhancement to be determined from measured SERS spectra. The relative chemical enhancement is found to be in reasonable agreement with the results of density functional theory modeling.

Vibrational frequencies and relative Raman scattering cross sections of SERS spectra are generally different from those measured in neat samples [1, 14]. As a consequence, additional bands may become observable in SERS spectra or existing bands may be attenuated beyond the detection limit. These modifications of Raman spectra complicate molecular identification, especially in mixtures. Within a classical electromagnetic model, surface selection rules are utilized to describe enhancement or attenuation of Raman scattering cross sections on flat [15, 16] or spherical [17, 18] surfaces. These rules have been used to identify the orientation of molecules on SERS substrates [19-21]. An alternative hypothesis [14] attributes spectral modifications in Raman spectra to charge-transfer excitations between molecules and metallic surfaces [9], which are not dependent on molecular orientation in a simple fashion.

Recent experimental developments in top-down and bottom-up nanofabrication allow control of electromagnetic near-fields in SERS substrates [22, 23]. These techniques open an opportunity for a deeper understanding of different contributions to Raman spectra enhancement.

In this Letter, we present experimental and theoretical studies of SERS substrates with nanoengineered plasmonic resonances, for which the modification of Raman spectra can be factored into electromagnetic and chemical components. While the first component characterizes the substrate, the second contribution contains additional information about molecular binding. We present evidence that changes in molecular electronic structure of the analyte may be more important for description of modification effects in SERS spectra than the effect of molecular orientation relative to the polarization of plasmon excitations in the metal substrate [18, 20]. Accordingly, the chemical contribution computed using a model geometry of a molecule adsorbed on small metal clusters is in a good agreement with the measured results.

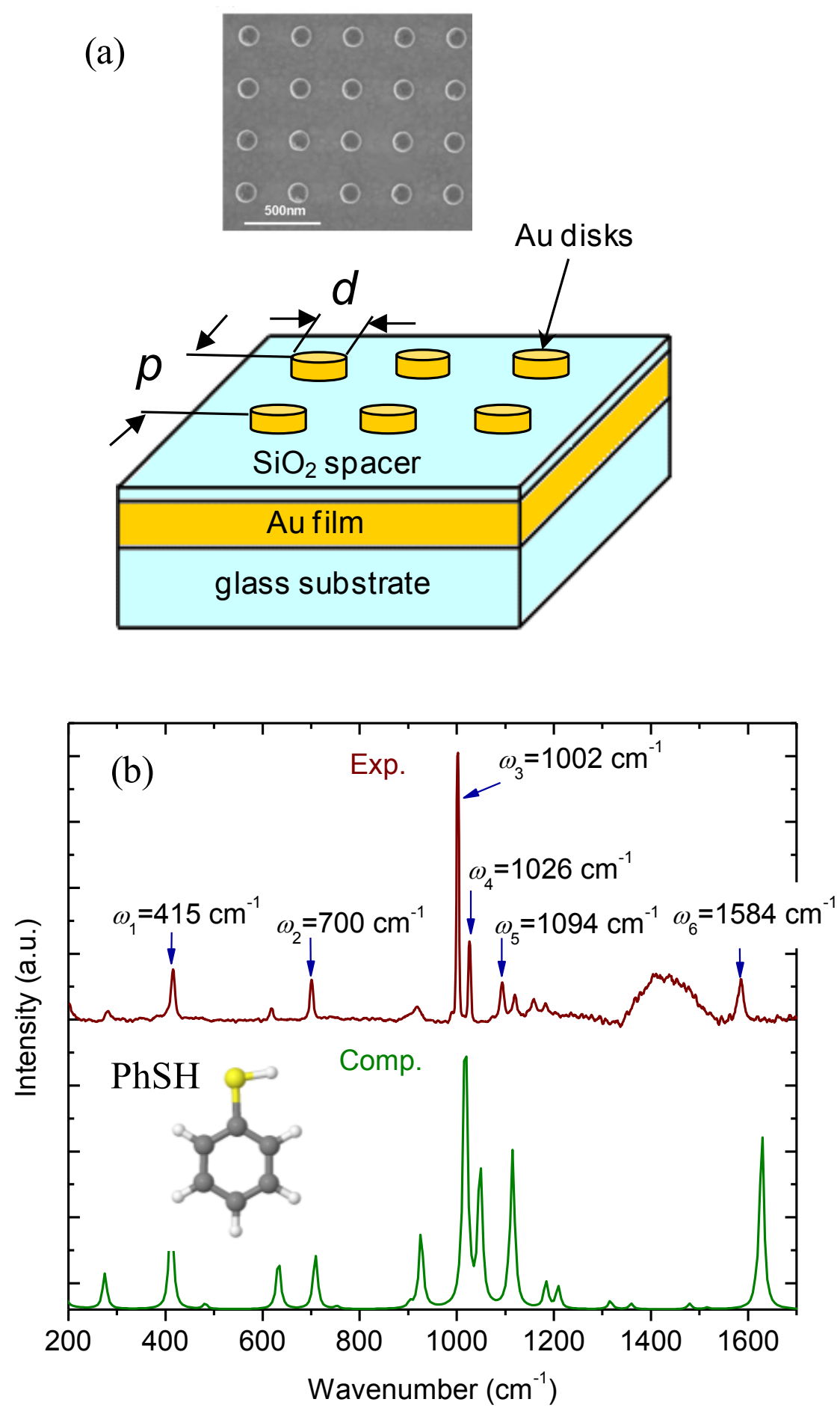

**Figure 1.** A scanning electron micrograph image and a schematic picture of a plasmon substrate utilized in the study (a); a measured and a computed Raman spectra of benzenethiol (b). The main Raman-active vibrational modes, which are also strongly enhanced in SERS, are shown.

We studied Raman spectra of benzenethiol adsorbed on nanoengineered substrates, see Fig 1. SERS spectra and extinction spectra at visible and near-infrared wavelengths were measured for 16 different SERS substrates. The substrates, Fig. 1(a), consisted of square lattices of gold metal disks with lattice periods $p$ = 350, 500, 770, and 780 nm. The heights of the disks were $h$ = 40 nm, while their diameters were varied in the range $d$ = 125-140 nm. In each substrate, the disks were formed on a $SiO_2$ spacer on a 100 nm thick gold film. The coupling between localized surface plasmons (LSPs) on the nanodisks and surface plasmon polaritons (SPPs) at the interface between the $SiO_2$ spacer and the gold film were controlled by varying the nanodisk diameter and the lattice period. Extinction spectra of the nanoengineered SERS substrates varied from a single-peak to a double-resonance shape, depending on the geometric parameters of the substrates, as shown in Fig 2(a). The observed changes in extinction spectra can be explained by hybridization of the LSP and SPPs [24]. In the following discussion, we divide experimental results into four groups, according to the lattice period of gold disk arrays.

The main Raman-active vibrational modes $\omega_1 = 415$ cm$^{-1}$, $\omega_2 = 700$ cm$^{-1}$, $\omega_3 = 1002$ cm$^{-1}$, $\omega_4 = 1026$ cm$^{-1}$, $\omega_5 = 1094$ cm$^{-1}$ and $\omega_6 = 1584$ cm$^{-1}$ are shown in Fig. 1(b). We characterized the modification of the spectra by shifts of these modes from their positions in the neat spectrum $\Delta\omega_n$ and by their relative enhancement $RE(\omega_n)$. The mode at $\omega_3 = 1002$ cm$^{-1}$ was used as a reference mode, as it is the strongest vibrational mode in Raman spectra of neat benzenethiol. The total relative enhancement factor was introduced for each mode $\omega_n$ as

$$RE_{\text{total}}(\omega_n) = \frac{I_{\text{SERS}}(\omega_n)}{I_{\text{SERS}}(\omega_3)} \frac{I_{\text{R}}(\omega_3)}{I_{\text{R}}(\omega_n)}, \tag{1}$$

where $I_{\text{SERS}}(\omega_n)$ and $I_{\text{R}}(\omega_n)$ are the intensities of the vibrational mode $\omega_n$ in SERS and conventional Raman spectra, respectively [25]. By this definition, therefore, the relative enhancement of the $\omega_3 = 1002$ cm$^{-1}$ mode is unity. Unlike absolute enhancement factors [26], relative enhancement factors are independent of the analyte concentration. For substrate-averaged characteristics, we neglect the correlations between molecular orientation and spatial dependence of local electromagnetic fields, and write the total relative enhancement factor as a product of electromagnetic and chemical enhancement factors,

$$RE_{\text{total}}(\omega_n) = RE_{\text{chem}}(\omega_n) \cdot RE_{\text{e.m.}}(\omega_n), \tag{2}$$

where the relative electromagnetic enhancement factor, $RE_{\text{e.m.}}(\omega_n)$, is approximated by the ratio of extinction intensities of the SERS substrate at the Stokes frequencies $\omega_n$ and $\omega_3$. This assumption is supported by recent wavelength-scanned SERS studies [22], which show that enhancement of the Raman scattering cross section follows the substrate extinction profile according to the conventional $|E|^4$ rule [27].

An extinction cross section of a metal nanoparticle is the sum of absorption and scattering cross sections and its spectral lineshape is therefore in general different from that of the near-field intensity enhancement on the metal surface [28-30]. For instance, different peaks in the extinction spectra can originate from different spatial positions on a substrate, some plasmon resonances can be not coupled to the far field [28], and the near- and the far-field peaks can be shifted with respect to each other [30]. On the other hand, as shown in Fig. 2(b), for our substrates, extinction spectra computed with the finite-difference time-domain (FDTD) method are in good agreement with the frequency dependence of the near field computed at a hot-spot point on the circumference of the bottom surface of the gold disk [23].

This result provides a basis for the above approximation. Moreover, the use of relative enhancements instead of absolute values makes the approximation less sensitive to dissimilarities of the spectra. However, for a more detailed analysis, an experimental measurement of near-field enhancement, for example using two photon photoluminescence [31], may be required.

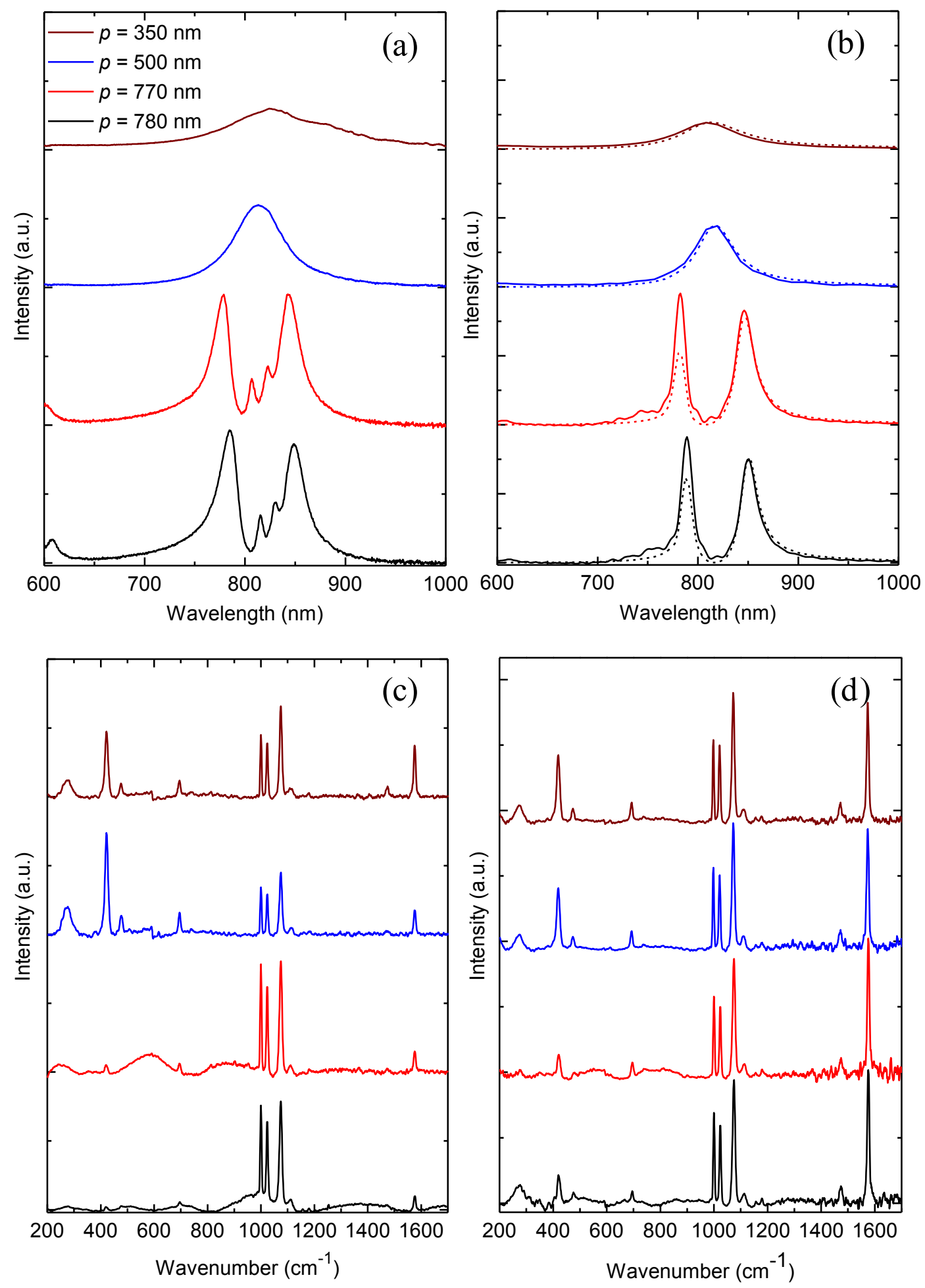

**Figure 2.** Examples of the measured extinction spectra (a), the extinction spectra (solid lines) and the near-field spectra (dashed lines) computed with FDTD method (b), SERS spectra (c), and SERS spectra corrected for electromagnetic deformation (d) measured on 4 different substrates with the lattice period $p$ = 350, 500, 770 and 780 nm and the disk diameter $d$ = 130 nm. The spectra are vertically displaced for clarity of presentation. The Raman spectra are corrected for the fluorescent background.

Variations in the plasmon resonance modify the electromagnetic component in Eq. (2), leaving the chemical modification unchanged [32]. In Figs. 2 (c) and (d) we show examples of the measured SERS spectra of benzenethiol for four substrates, each having a different period, and the SERS spectra corrected for electromagnetic modification using Eq. (2). The correction process consists of dividing the measured Raman spectra by the extinction cross sections at the excitation and the Stokes frequencies. In the unprocessed spectra the relative intensities of Raman lines vary by an order of magnitude. Variations in the peak heights of the spectra corrected for the inhomogeneous electromagnetic enhancement are less than 15% amongst all 16 spectra, with the exception of the $\omega_1 = 415$ cm$^{-1}$ mode. The relative intensity of this mode does not fluctuate much for the substrates within a group, but it differs by a factor of up to 2.5 if spectra from different groups are compared [33]. The latter difference is attributed to the fact that the intensity of the $\omega_1$ mode may be underestimated for the substrates with plasmon double-resonance structures due to a considerable dissimilarity between the near- and far-fields. The $\omega_1$ mode occurs between the two plasmon peaks, a region where even a minor difference between the far- and the near-fields can substantially affect the results. Differences between the corrected Raman spectra, see Fig 2(d), and spectra of the neat analyte, see Fig. 1(b), are attributed to chemical effects.

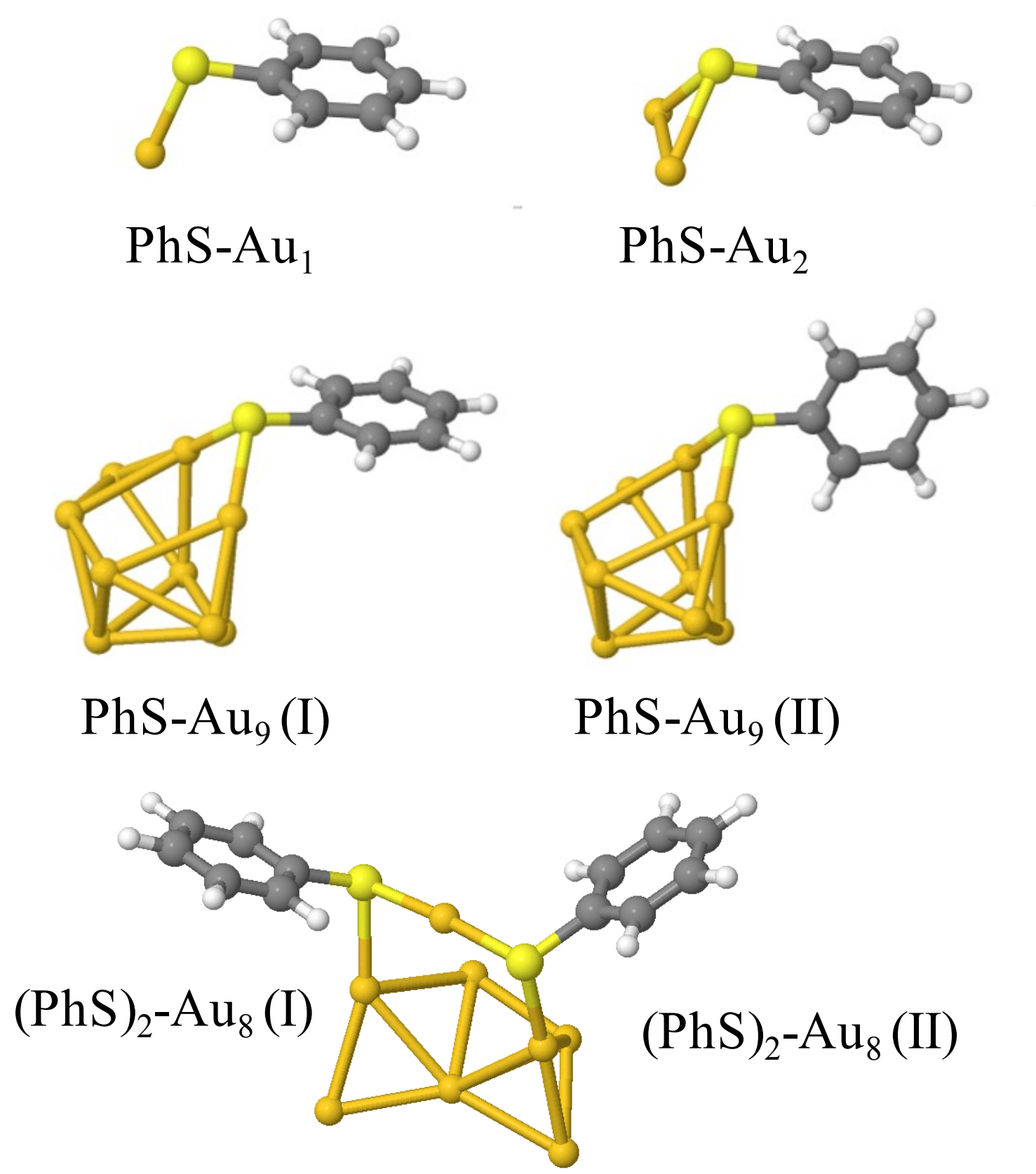

**Figure 3.** Computed PhS–Au$_n$ structures used to characterize the chemical modification of the Raman spectra.

To characterize the chemical modification of the spectra, we performed time-dependent density functional calculations of Raman spectra of isolated benzenethiol and five representative PhS–Au$_n$ complexes shown in Fig. 3. By using small metal-molecular complexes, we assumed that chemical effects in SERS are strongly localized near binding sites [12]. This model can account for atomic scale imperfections on a metal surface and is quite common in the theory of surface catalysis [35, 36]. The two structures PhS–$Au_1$ and PhS–$Au_2$ represent the simplest configurations of benzenethiol bound to Au. Two larger structures PhS–$Au_9$ (I) and (II) show more pronounced metallic properties, while the structure $(PhS)_2$–$Au_8$ represents a "staple" binding motif [34]. For a quantitative analysis of the chemical effect one needs to explore the large space of different cluster geometries. Here we selected several structures with a small number of metal atoms showing binding motifs found in other studies [12, 34]. We assume that the representative clusters provide enough degrees of freedom to mimic accurately the environment adsorbate molecules might meet near a surface. The geometries shown in Fig. 3 were optimized, which further restricted the allowed binding configurations. All computations were performed using Turbomole quantum chemistry package, version 5.10 [37, 38].

Rigorous calculations of Raman scattering cross sections require integration of the molecular response over all possible orientations relative to the surface. This procedure is complicated by the lack of information about atomic-scale roughness features, which have strong influence on surface selection rules [18]. Thus, we introduced two types of relative enhancement factors that characterize the geometrical effect. The orientation-averaged relative enhancement factor, $RE^{\mathrm{A}}_{\mathrm{chem}}(\omega_n)$, was defined as the ratio of differential scattering cross-sections

$$RE^{\mathrm{A}}_{\mathrm{chem}}(\omega_n) = \frac{d\sigma_{\mathrm{C}}(\omega_n)}{d\sigma_{\mathrm{C}}(\omega_3)} \frac{d\sigma_{\mathrm{PhSH}}(\omega_3)}{d\sigma_{\mathrm{PhSH}}(\omega_n)}, \tag{3}$$

where the index "C" denotes a particular PhS–$Au_n$ complex, and free-space averaging over all possible orientations of the structures relative to the probe field was applied [39]. Equation (3) accounts only for local effects on the length scale of the metal cluster. It completely neglects geometrical effects of fixed molecular orientation due to chemical attachment [18]. In the trivial case of isolated benzenethiol, the orientation-averaged relative chemical enhancement is equal to 1 for all modes. The fixed orientation estimate of the chemical modification was obtained by aligning all structures such that the connecting line between the sulfur atom and the nearest point of the metal cluster surface is parallel to the *z*-axis. Assuming that localized plasmon excitation is polarized along the *z*-axis and that only the $\alpha_{zz}$ component of the molecule Raman tensor contributes to the Raman response of a given mode $\omega_n$, the fixed orientation estimate for relative enhancement factors, $RE^{\mathrm{F}}_{\mathrm{chem}}(\omega_n)$, can be written as

$$RE^{\mathrm{F}}_{\mathrm{chem}}(\omega_n) = \frac{(\omega_{\mathrm{ex}} - \omega_n)^4 \omega_3}{(\omega_{\mathrm{ex}} - \omega_3)^4 \omega_n} \left( \frac{\alpha^{\mathrm{C}}_{zz}(\omega_n)}{\alpha^{\mathrm{C}}_{zz}(\omega_3)} \right)^2 \frac{d\sigma_{\mathrm{PhSH}}(\omega_3)}{d\sigma_{\mathrm{PhSH}}(\omega_n)}, \tag{4}$$

where the frequency dependent prefactor on the right hand side comes from the definition of differential cross sections, $\omega_{\mathrm{ex}}$ is the excitation frequency and the index C denotes a specific structure. The Raman response from an isolated benzenethiol computed with Eq. (4) neglects all effects related to formation of metal-molecular complexes.

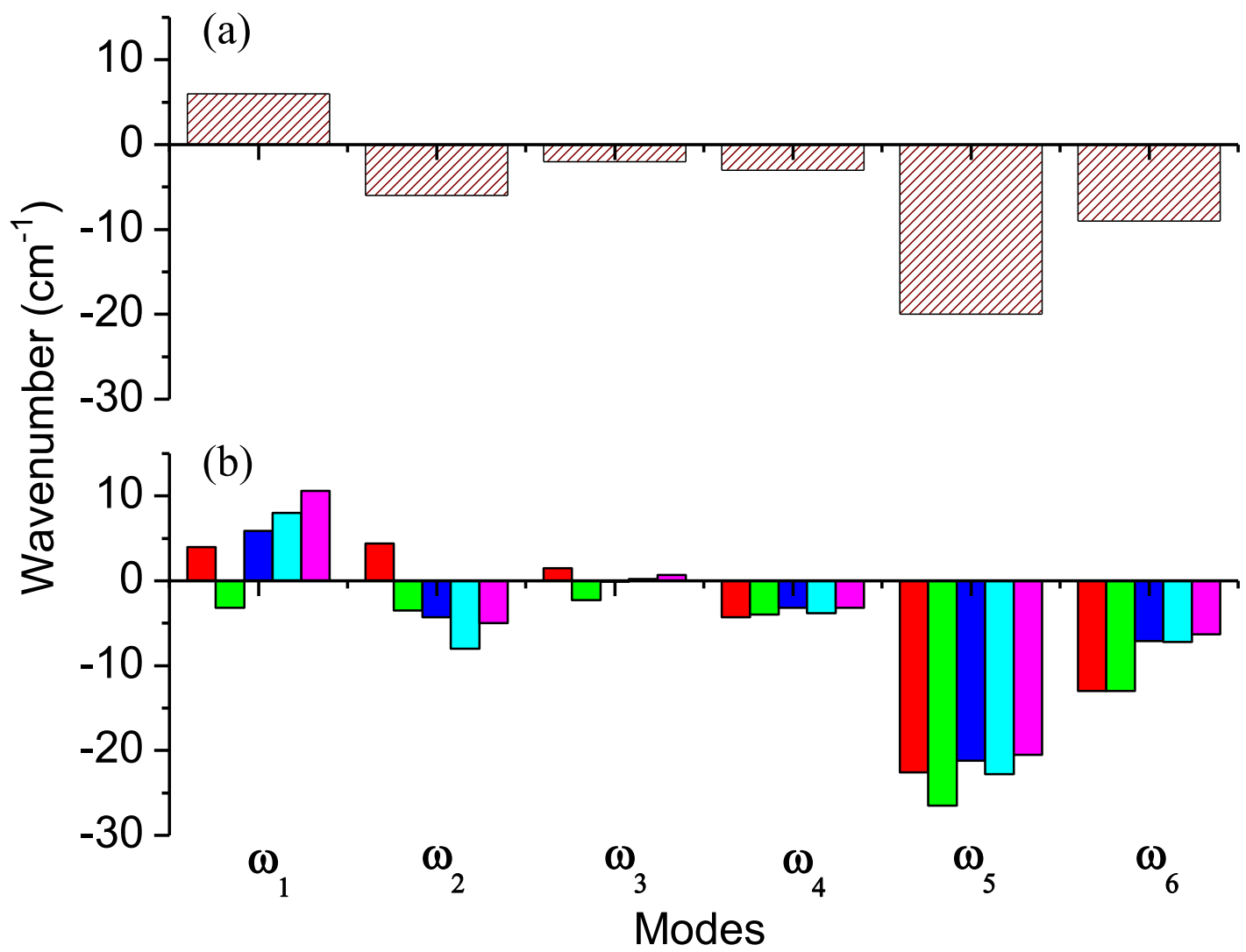

**Figure 4.** The shifts of 6 Raman active modes ($\omega_1 = 415$ cm$^{-1}$, $\omega_2 = 700$ cm$^{-1}$, $\omega_3 = 1002$ cm$^{-1}$, $\omega_4 = 1026$ cm$^{-1}$, $\omega_5 = 1094$ cm$^{-1}$, and $\omega_6 = 1584$ cm$^{-1}$) of benzenethiol adsorbed on nanoengineered substrates as compared to the computed shifts for different PhS–$Au_n$ complexes: (a) experimental data, the shifts are the same for all substrates; (b) computed shifts, red – PhS–$Au_1$, green – PhS–$Au_2$, blue – PhS–$Au_9$ (I), cyan – PhS–$Au_9$ (II), magenta – $(PhS)_2$–$Au_8$.

The computed spectrum of an isolated benzenethiol molecule, Fig. 1b, shows all the Raman modes observed in the experiment. Moreover, the frequency shifts and the relative enhancements used in the analysis are weakly sensitive to the absolute differences between the spectra. The computed frequency shifts, Fig. 4, and the relative chemical enhancement factors, Fig 5, are in a reasonable agreement with the measured results for all simulated metal-molecular complexes. In contrast, the fixed orientation of the molecule relative to the metal surface is not sufficient to reproduce strong modifications in Raman spectra of molecules attached to metal surfaces. For example, the orientation-fixed relative enhancement factor of the $\omega_5 = 1094$ cm$^{-1}$ mode for an isolated benzenethiol is equal to 1.1, which should be compared to over 10 times enhancement observed experimentally on plasmonic substrates. Variations in the molecular orientation relative to the polarization of a probe field do not account for the strong enhancement either. Among the modes $\omega_1 - \omega_6$, the strongest relative enhancements are observed for the modes that show the strongest vibrational frequency shifts.

To elucidate the role of chemical binding we analyze the change of the molecular structure due to adsorption. In the computed metal-molecular complexes, some electron density is transferred from the metal cluster to the molecule. We estimate by natural population analysis [40] that about 0.2 *e* are transferred from the cluster to the sulfur atom. For the PhS–$Au_1$ and PhS–$Au_2$ complexes the amount of charge transferred to the sulfur atom is about 10-30% lower, which can be attributed to larger ionization potentials of smaller metal clusters. For all computed structures, the C–S bond is elongated by 0.5-1% and the adjacent C–C bonds are elongated by 0.2-0.3% as compared to isolated benzenethiol. These modifications are associated with force constants of the corresponding bonds.

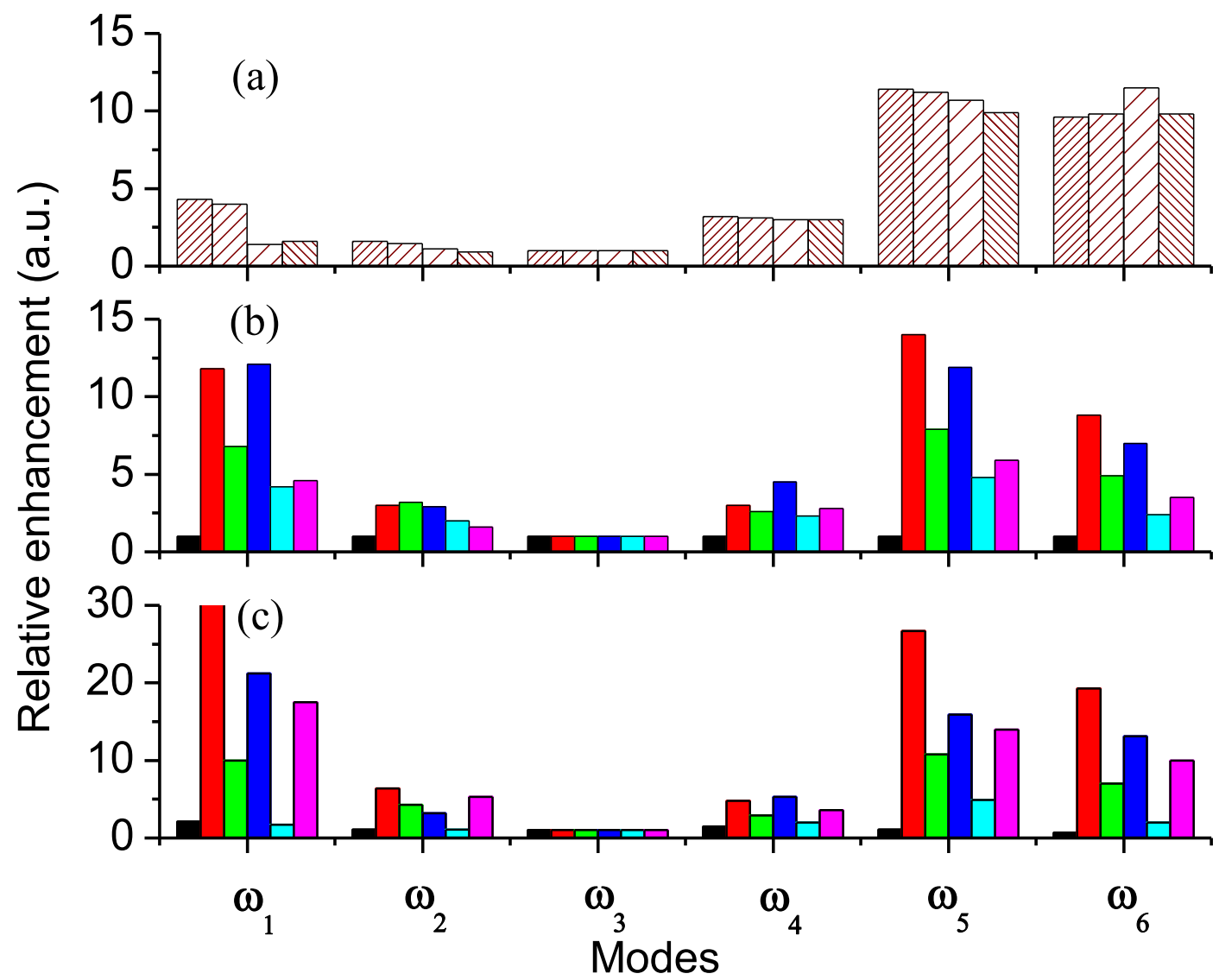

**Figure 5.** The chemical relative enhancement factors obtained by correcting the measured spectra for the electromagnetic modification (a) compared to orientation-averaged estimates, Eq. (3), (b) and orientation-fixed estimates, Eq. (4), (c) for isolated benzenethiol and 5 PhS–$Au_n$ structures from Fig. 3. It should be noted that the vertical scale for the orientation-fixed estimates is different by the factor of 2 from two other plots. The measured enhancement factors are averaged over the substrates within the same group. The color coding is as follows: black – PhSH, red – PhS–$Au_1$, green – PhS–$Au_2$, blue – PhS–$Au_9$ (I), cyan – PhS–$Au_9$ (II), magenta – $(PhS)_2$–$Au_8$.

All studied vibrational modes, $\omega_1 - \omega_6$, involve multiple bond stretches. The lowest-frequency mode, $\omega_1 = 415$ cm$^{-1}$, in metal molecular-complexes acquires a substantial admixture of Au–S bond stretching component upon adsorption. Interaction with the low-frequency Au–S stretching mode may explain the blue-shift of the $\omega_1$ mode. Similar blue-shifts were observed for several out-of-plane modes of the benzene ring with rather small Raman intensities. These modes also involve vibrations of the sulfur atom with respect to the Au cluster. The largest discrepancy with the experimental data found for $\omega_1 = 415$ cm$^{-1}$ mode can also be explained by its sensitivity to a specific geometry of the local metal environment. The red-shifts of the modes $\omega_2 - \omega_5$ are associated with the reduction of the C–S and C–C bond strength. These vibrational modes do not overlap with the metal cluster. The shift of the $\omega_5$ mode and the corresponding intensity change can be explained by the C–S and the C–C stretch involving the carbon atom adjacent to sulfur. The presence of an additional vibrational mode at 1473 $cm^{-1}$ is also consistent with the hypothesis of a bonding effect. In the computed spectra, this mode is very weak for isolated benzenethiol and is strongly enhanced in metal-molecular complexes. It involves the C–S and the C–C bond stretches, similarly to $\omega_5 = 1094$ $cm^{-1}$. These results support the findings of Refs. [11, 41], in which vibrational modes of pyridine attached to a silver adatom were analyzed. In our case the changes are more delocalized due to a stronger binding and involve all molecular vibrations. These results do not provide a simple unified picture which suggests that the details of the changes in electronic structure are only obtainable from specific computations.

Similar to the Raman signal enhancement [11] we can assign two distinct contributions to the spectral modifications: the static contribution associated with increase of polarizability derivatives due to change in ground-state electronic structure upon attachment to a metal surface, and the dynamic contribution associated with the formation of metal-molecular states [12] or charge-transfer excitations [42]. As an illustration, in Fig. 6, the computed orientation-averaged relative enhancements for the PhS–$Au_9$ (I) complex are shown as functions of the excitation energy. This plot is an analogue of a Raman excitation profile [12] introduced for relative enhancements. The frequency dependence is given by sensitivity of molecular vibrational modes to the electronic transitions. In the particular example the lowest electronic transition involving the molecule is above the shown spectral range. The static component of spectral modification can be approximated by relative enhancements of vibrational modes computed about 1500 nm. At this wavelength the frequency dependence of the relative enhancements is essentially flat. The static component accounts for more than 50% of relative enhancement factors taken at 783 nm. A stronger frequency dependence of the $\omega_1$ and $\omega_5$ modes is natural, because of their proximity to the bonding site. The actual energies of the metal-molecular states may differ from the ones computed with metal-molecular complexes. Therefore, the analysis we provide is applicable at least for a non-resonant excitation, where excited electronic states are accounted for in an integrated fashion. Thus, it would be interesting to analyze the modification of Raman spectra in molecules that may have a distinct contribution of a charge-transfer excitations, such as 4-aminobenzenethiol [43].

*In conclusion*, we have characterized changes in Raman spectra of benzenthiol molecules adsorbed on nanoengineered plasmonic substrates. Two major contributions have been identified: the variation in electromagnetic enhancement with frequency and the chemical effect. Modifications to the intensities of the lines of Raman spectra can be broken down into these two contributions for substrates with plasmon resonances profiles varying from single-peaked localized plasmon structures to double-peaked hybrid localized-propagating plasmon structures. The chemical modification is associated with changes of electronic structure of the adsorbed molecule. We have shown that access to the extinction spectra of nanoengineered plasmonic substrates in combination with calculations of small cluster models of metal-molecular complexes provides a readily accessible tool for a qualitative characterization of chemical modification effect. Possible applications of the methodology introduced in this Letter include the development of computational databases of probable binding motifs of analytes to small coinage metal clusters, which combined with electromagnetic substrates could allow for the automated identification

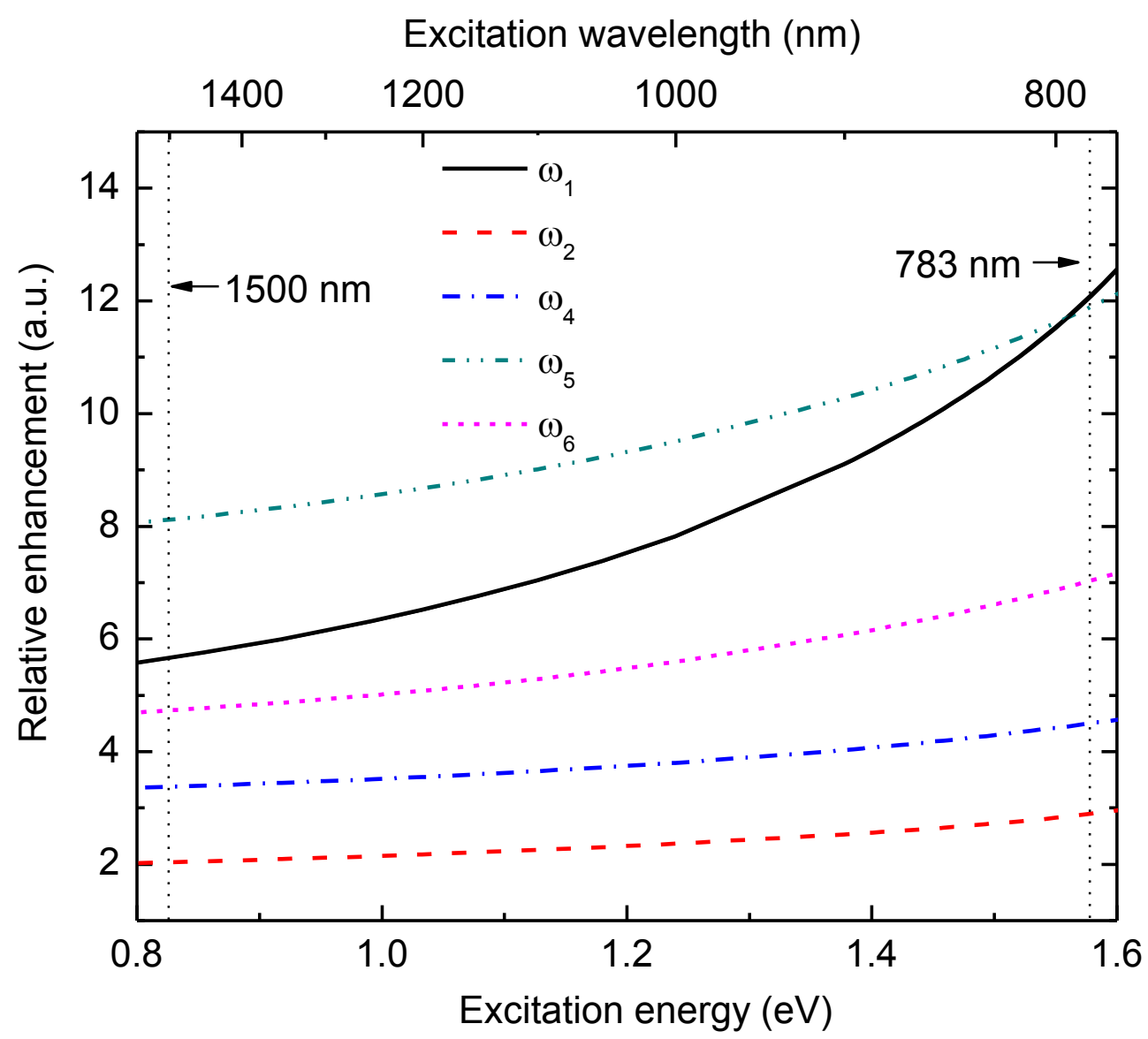

**Figure 6.** Computed excitation profiles for the orientation-averaged relative enhancement of 5 vibrational modes $\omega_1 = 415$ cm$^{-1}$, $\omega_2 = 700$ cm$^{-1}$, $\omega_4 = 1026$ cm$^{-1}$, $\omega_5 = 1094$ cm$^{-1}$, and $\omega_6 = 1584$ cm$^{-1}$ in the PhS–$Au_9$ (I) complex. The reference mode is $\omega_3 = 1002$ cm$^{-1}$. The 1500 nm excitation limit represents the static contribution in the chemical modification effect.

ACKNOWLEDGMENTS: This work was supported by the Defense Advanced Research Project Agency under Contract No FA9550-08-1-0285 and the Defense Threat Reduction Agency under Contract No HDTRA1-10-1-0046. We would like to thank the High Performance Technical Computing Center at the Faculty of Arts and Sciences of Harvard University and the National Nanotechnology Infrastructure Network Computation project for providing the computational resources.

SUPPORTING INFORMATION AVAILABLE: experimental and computational details, population analysis of the computed structures, computed Raman spectra, Raman tensors, relative enhancement factors and frequency shifts. This material is available free of charge via the Internet at http://pubs.acs.org